\documentclass[%
 reprint,
 amsmath,amssymb,
 aps,
floatfix
]{revtex4-2}
\usepackage{mathtools}
\usepackage{braket}
\usepackage{graphicx}
\usepackage{dcolumn}
\usepackage{bm}
\usepackage{float}
\usepackage{color}
\usepackage{soul}

\begin{document}

\preprint{DMRG_Sampled_Variance}

\title{ Unusual energy spectra of matrix product states}

\author{J. Maxwell Silvester\textsuperscript{1}}
\author{Giuseppe Carleo\textsuperscript{2}}
\author{Steven R. White\textsuperscript{1}}
\affiliation{
    \textsuperscript{1}Department of Physics and Astronomy, University of California, Irvine, California 92667, USA
}
\affiliation{
    \textsuperscript{2}Institute of Physics, École Polytechnique Fédérale de Lausanne (EPFL), CH-1015 Lausanne, Switzerland
}

\date{\today}

\begin{abstract}
In approximate ground states obtained from imaginary-time evolution, the spectrum of the state---its decomposition into exact energy eigenstates---falls off exponentially with the energy. 
Here we consider the energy spectra of approximate matrix product ground states, such as those obtained with the density matrix renormalization group.  Despite the high accuracy of these states, contributions to the spectra are roughly constant out to surprisingly high energy, with
an increase in the bond dimension reducing the amplitude but not the extent of these high-energy tails.
The unusual spectra appear to be a general feature of compressed wavefunctions, independent of boundary or dimensionality, and are also observed in neural network wavefunctions. The unusual spectra can have a strong effect on sampling-based methods, yielding large fluctuations. The energy variance, which can be used to extrapolate observables to eliminate truncation error, is subject to these large fluctuations when sampled. Nevertheless, we devise a sampling-based variance approach which gives excellent and efficient extrapolations. 

\end{abstract}

\maketitle

Numerical simulations of a variety of sorts have become essential for the study of quantum many-body systems.  
Given an approximate ground state, $\ket{\psi}$, we consider its decomposition into exact energy eigenstates, i.e., its energy spectrum, 
\begin{equation}\label{eq:200}
\ket{\psi} = \sum_n c_n \ket{n} , 
\end{equation}
where $\{\ket{n}\}$ are the eigenstates of the Hamiltonian with corresponding energies $\{E_n\}$.
For a number of numerical approaches based on imaginary-time evolution, e.g. quantum Monte Carlo, the coefficients $c_n^2$ fall off exponentially with the gap $E_n - E_0$, where $E_0$ is the ground-state energy~\cite{Becca_Sorella_2017, shi2018variational}. 
This exponential fall-off provides strong guarantees on convergence rates, particularly where there are no near-degeneracies. 
Because the density matrix renormalization group (DMRG) can achieve very accurate ground-state energies, one might assume that the high-energy coefficients of the corresponding matrix product state (MPS) would also fall off rapidly as the energy gap increases~\cite{white1992density,schollwock2011density,Stoudenmire_2012, PhysRevB.79.220504,PhysRevLett.109.067201,doi:10.1126/science.1201080}.  
Surprisingly, we find that the high-energy coefficients do not decrease exponentially with the gap.
In fact, for a substantial energy range they do not decrease much at all.  

\begin{figure}[h!]
\includegraphics[width=\linewidth]{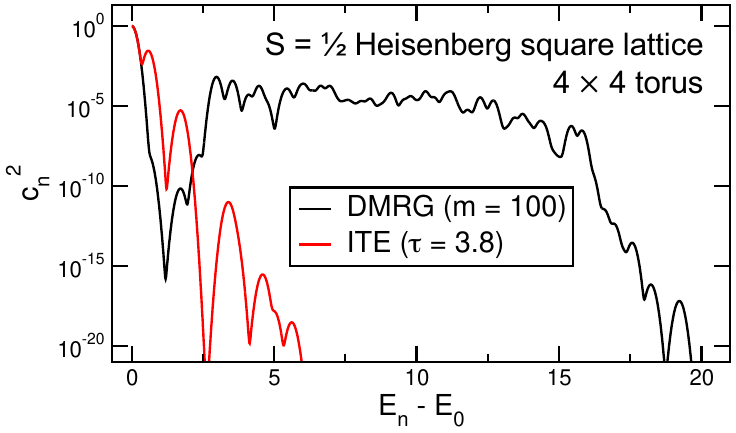}
\caption{\label{fig01}  Energy spectra of approximate ground states from DMRG and imaginary-time evolution (ITE) for a small cluster with fully periodic boundary conditions, broadened by  Gaussians with width $0.1$.  The states are matched to have the same energy, with $E - E_0 = 0.016$, but the energy variance $\sigma_{\hat{H}}^2$ is much larger for DMRG, $0.119$ versus  $0.0092$.}
\end{figure}

In Fig.~\ref{fig01} we show spectra for a spin $S = 1/2$ Heisenberg model on a $4 \times 4$ square lattice cluster.  Since producing the spectrum of a DMRG state requires a full diagonalization of the Hamiltonian, it is only available on small test systems.  
This model is described by the Hamiltonian
\begin{equation}\label{eq:15}
H = J \sum_{\braket{i,j}} \vec{S}_i \cdot \vec{S}_j,
\end{equation}
where the sum ranges over nearest-neighbor pairs. We set $J = 1$ throughout, corresponding to antiferromagnic coupling. For this system, we compare DMRG results to those from an exact imaginary-time evolution (ITE) starting with a Néel state. 
The imaginary-time duration, $\tau$, is chosen to make the two states have approximately the same energy.  
A clear exponential tail is seen for the ITE spectrum, as expected.  
In contrast, the DMRG spectrum has very little weight in the first few excited states and a tail that is roughly constant out to surprisingly high energies.  Note that we choose fully periodic boundary conditions, for which DMRG requires larger bond dimension to achieve accurate results, in order to emulate a DMRG run on a larger system while still being able to fully diagonalize the Hamiltonian. 

We obtain similar results for different lattice models and geometries, with states obtained both from DMRG and from MPS compression of an exact ground state \cite{SM}. In this Letter, we present these results and explore important implications of such spectra for sampling-based methods.

\begin{figure}[h!]
\includegraphics[width=\linewidth]{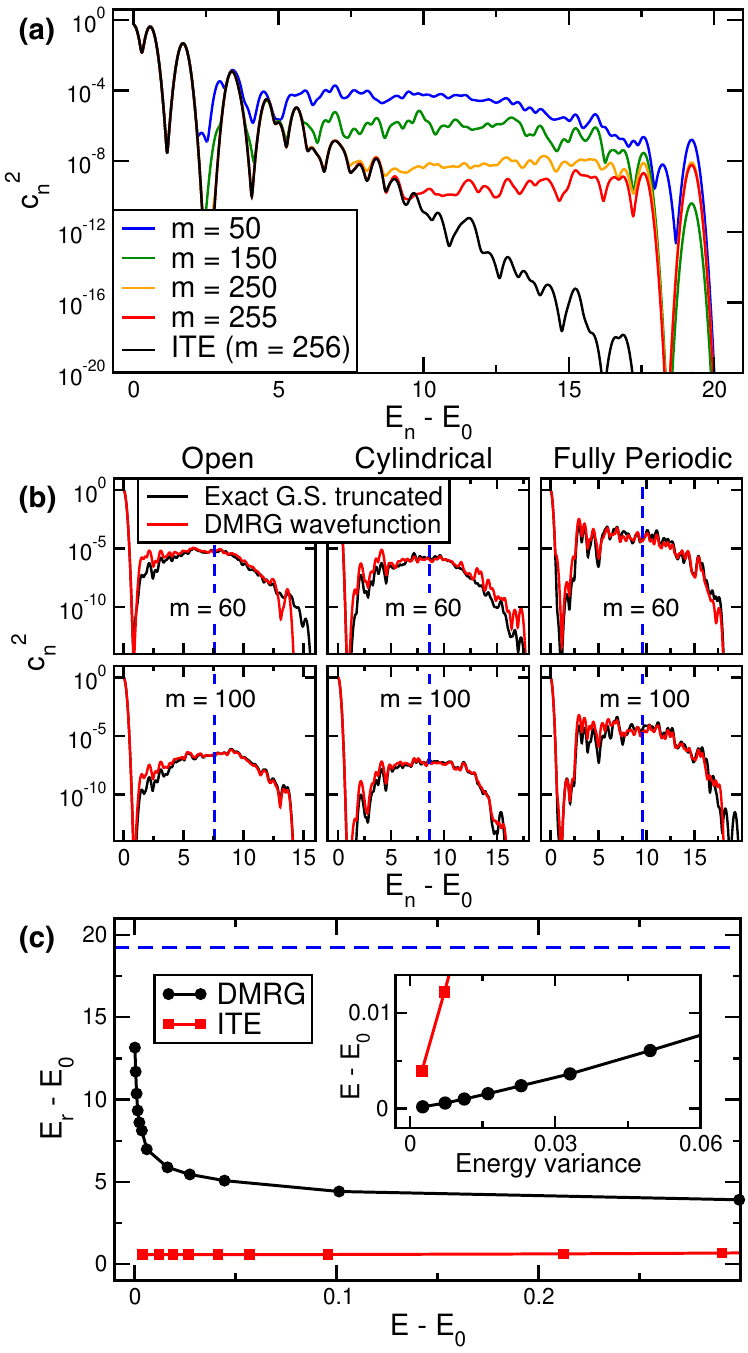}
\caption{\label{fig02} Results for a $4 \times 4$,  $S = 1/2$ Heisenberg square lattice with various boundary conditions. (a) For fully-periodic boundary conditions (PBC), the effect of truncating an ITE state ($\tau = 1$) when it is represented as an MPS. Energy spectra obtained via exact diagonalization (ED) and broadened with Gaussians. (b) Comparing the states obtained via truncating the exact ground states to bond dimension $m$ vs the the states obtained via running DMRG at the same bond dimension.  Again, results are obtained via ED and broadened. The vertical dashed lines indicates half the difference between the maximum and minimum eigenenergies of the system, i.e., half the bandwidth. (c) For PBC, the energy of the residue vs the energy error of solutions from both DMRG and ITE.  The dashed line indicates the full bandwidth of system. On the inset, the energy error of the state vs the energy variance.  Points on the main plot and inset indicate distinct bond dimensions or imaginary-time steps for the DMRG and ITE states, respectively.  }
\end{figure}

\emph{Unusual spectra}---The unusual energy spectra are due to the compression of the state into an MPS,
rather than some aspect of the DMRG energy optimization.  
This is shown in Fig.~\ref{fig02}(a), where we compress an ITE state into an MPS. 
The spectrum flattens at an energy gap determined by the bond dimension, $m$. 
Note that for this small cluster, the maximum bond dimension for any state to be represented exactly is $m = 256$,  the Hilbert space dimension of half the system. 
It is remarkable that the flat tail is apparent even with a truncation to $m = 255$. 
While some details of the results of Fig.~\ref{fig02}(a) are specific to the fully periodic boundaries, the qualitative behavior is independent of boundary conditions.  
This is illustrated in Fig.~\ref{fig02}(b), where we compare DMRG spectra to spectra obtained from truncating the exact ground-state to the same bond dimension as the corresponding DMRG state.  
Regardless of bond dimension or boundary conditions, the spectra show similar qualitative features, extending to high energy.  
Moreover, in each case the DMRG spectrum matches closely the spectrum from truncation of the exact solution, confirming that indeed the spectra are a result of the compressed MPS representation of the wavefunction.

It is also interesting to consider the average energy of a DMRG state when the exact ground state is projected out, i.e., the energy of the excited part of the solution. 
Thus, we decompose the approximate ground state as 
\begin{equation}\label{eq:5}
\ket{\psi} = \sum_{n} c_n \ket{n} = \cos\theta \ket{0} + \sin\theta \ket{r},
\end{equation}
where the \emph{residue}, $\ket{r}$, is normalized.  
In Fig.~\ref{fig02}(c) we show the average energy of the residue, i.e., $E_r = \braket{r|\hat{H}|r}$, versus energy error as we vary the bond dimension in a DMRG calculation.  
Remarkably, $E_r$ increases sharply as the bond dimension is increased, eventually reaching about $2/3$ of the total energy bandwidth (dashed blue line).  
Of course, this increase in the residue's energy is more than compensated for by a decrease in $\theta$, so that the $E - E_0$ still approaches zero.  
In contrast, for the ITE state, with increasing $\tau$, $E_r$ simply approaches $E_1 = \braket{1|\hat{H}|1}$, the energy of the first excited state.

The striking dissimilarity between the residues of the DMRG and ITE states provides a way to probe the energy spectra of larger systems, where exact diagonalization is unavailable.  
Note that for the special case when $\ket{r} = \ket{1}$, i.e., when all of the error in the approximate ground-state is in the first excited state, the quantity
\begin{equation}\label{eq:7}
\Tilde{v} = (E - E_0)^2 \cos^2 \theta + (E - E_1)^2 \sin^2 \theta. 
\end{equation}   
is  equal to the wavefunction's energy variance, $\sigma_{\hat{H}}^2 =  \bra{\psi} (\hat{H} - E)^2 \ket{\psi}$. 
Importantly, calculating $\Tilde{v}$ requires only quantities that can be easily estimated within DMRG, such as the lowest energy gap. 
The ratio $\sigma_{\hat{H}}^2/\Tilde{v}$ serves as a measure of the high-energy contributions to the wavefunction.  
Specifically, for ITE states, this ratio reduces to unity as the ground state is approached.
In contrast, for DMRG wavefunctions this ratio steadily increases as the MPS converges, reaching values of $10^2 - 10^4$ or higher. 
This prominent signal is seen across different geometries, boundary conditions, and models, as described in the SM~\cite{SM}, showing that  high-energy residues are a quite-general feature of MPS solutions.

Why do MPS spectra have these high-energy tails? 
To be specific, let us suppose that our approximate ground state has $m=100$, but that high accuracy representations of the ground and low-lying excited states requires $m=1000$ (a still modest size corresponding to states with area-law entanglement).  Note that the total number of parameters in an MPS of length $N$ is about $N m^2$. 
Composing the $m=100$ state exactly out of several of the $m=1000$ states would require a precise cancellation of about $10^6$ parameters per site into only $10^4$. If, instead, the number of states used to make the approximate ground state was comparable to the number of parameters in each state, the cancellation could easily occur.  The energy levels
at low energy are not nearly as dense as at the center of the spectrum, so obtaining enough states to match the number of parameters requires extending the spectrum substantially into the bulk. However, in the bulk, the states have volume law entanglement, and the number of parameters in each is exponential in the length. Thus, we need an exponentially large number of states, and the spectrum covers the entire energy range.  

Since this reasoning is not very specific to MPS states,  we expect similar behavior for other types of compressed wavefunctions.  In fact, for the same $4\times4$ torus we find that neural network states~\cite{carleo_solving_2017} have strikingly similar spectra to MPS solutions~\cite{SM}.

As shown on the inset of Fig.~\ref{fig02}(c), DMRG states have a high variance relative to the energy error, making comparisons to other numerical methods based on energy variance potentially misleading~\cite{varbench}. Moreover, computing the exact energy variance of a DMRG wavefunction is more expensive than a DMRG sweep by a factor of $w$, the bond dimension of the Hamiltonian expressed as a matrix product operator. For systems with long-range interactions or two-dimensional (2D) geometry we may have $w \geq 30$. Nevertheless, the near-linear behavior shown in the inset indicates that the variance may be an ideal candidate for extrapolating the energy to the  zero variance limit.  Recently, the \emph{two-site variance} has been introduced as a means to approximate the energy variance of the DMRG wavefunction at an affordable cost, enabling extrapolations when running single-site DMRG~\cite{TwoSiteVar}.  
However, the cost of computing the two-site variance is similar to that of a DMRG sweep, roughly doubling run-time.

\emph{Consequences for sampling}--- As an alternative approach to estimating the variance cheaply, we consider using MPS sampling, which also illustrates the strong effect that unusual MPS spectra have on sampling-based methods.  
Unlike in many quantum Monte Carlo algorithms, which construct a Markov chain with nonzero autocorrelation times, MPS states can be sampled with one perfectly independent sample per step. 
This perfect sampling technique is the starting point for both the minimally entangled typical thermal state algorithm and MPS-based variational Monte Carlo methods~\cite{MilesSampling,VidalSampling,sandvik2007variational,wouters2014projector}.  

A sample is a product state, $\ket{s}$, where the product is over basis states of each site. The probability of generating a sample is $P_s =  | \braket{\psi|s} |^2 $, and we have
\begin{equation}\label{eq:1}
\sigma^2_{\hat{H}} = \sum_{s
}  \braket{\psi|\hat{H} - E|s} \braket{s|\hat{H} - E|\psi},
\end{equation}
where $E = \braket{\psi|\hat{H}|\psi}$ is the expectation value of the energy and the sum ranges over all product states, which form a complete basis.
The \emph{support}, $\mathcal{S}$, is the set of basis vectors for which $P_s$ is nonzero.
For states in $\mathcal{S}$, we define the \emph{local energy}, $E_s^L = \frac{\braket{s|\hat{H}|\psi}}{\braket{s|\psi}}$; outside $\mathcal{S}$, the local energy is undefined.  
Then, splitting the sum in Eq.~(\ref{eq:1}) into two parts, we have
\begin{equation}\label{eq:2}
\sigma^2_{\hat{H}} = \Delta_s + \sum_{\ket{s} \in \mathcal{S}} 
P_s |E_s^L - E |^2 ,
\end{equation}
where the \emph{local energy bias}  is
\begin{equation}\label{eq:3}
\Delta_s = \sum_{\ket{s} \notin \mathcal{S}}  
|\braket{\psi|\hat{H} |s}|^2.
\end{equation}
Note that, alternatively, an unbiased estimator can be obtained by inserting the identity operator next to $(\hat H-E)^2$ in Eq.~(\ref{eq:1}), instead of between the two copies of $(\hat H-E)$. 
In that case the samples with $P_s=0$ do not contribute to the variance. The sampling bias $\Delta_s$ rigorously vanishes when $\ket{\psi}$ is exact, and in the systems we have studied, it is usually small.  We also find that the unbiased version of sampling is both more expensive and also prone to worse fluctuations, so we utilize the local energy sampling~\cite{SM}.

\begin{figure}[h!]
\includegraphics[width=\linewidth]{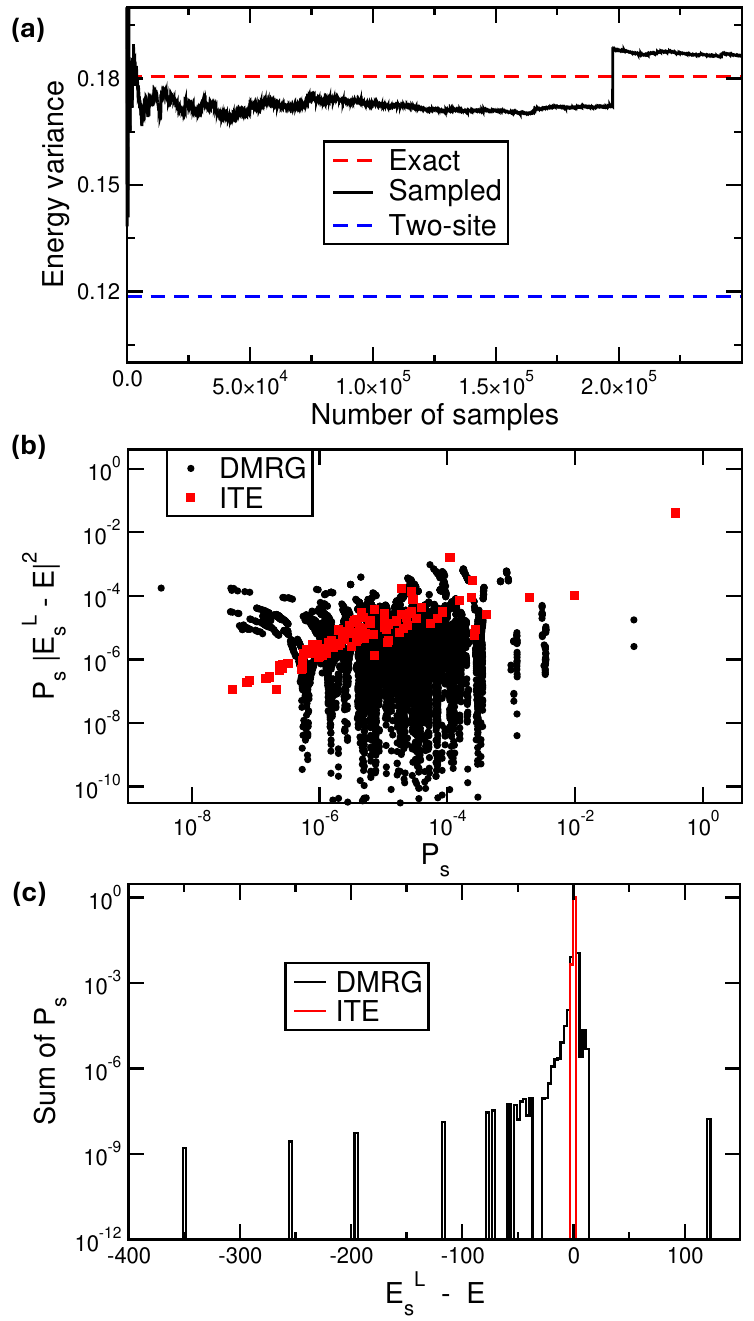}
\caption{\label{fig03} For the same $4\times4$ torus, (a) the running average of the sampled variance, the two-site variance, and the exact energy variance, all for the same DMRG state ($m = 80$). (b) The probability distribution of local energies for approximate ground-state solutions from DMRG and imaginary-time evolution (ITE), both with similar energy variance ($\sigma_{\hat{H}}^2 \approx 0.18$). The ITE state is initialized with Néel order.  As in Eq.~(\ref{eq:2}), the $y$-axis indicates each state's contribution to the energy variance. (c)  A histogram showing the probability of sampling a local energy in a certain range. }
\end{figure}

With $N_s$ samples, the sampling estimate of the energy variance is 
\begin{equation}\label{eq:4}
\sigma^2_{\hat{H}} - \Delta_s \approx \frac{1}{N_s} \sum_{i = 1}^{N_s} |E_{i}^L - E|^2.
\end{equation}
The cost of selecting a product state, $\ket{s}$, in a perfect measurement, is $\mathcal{O}(m^2d + md^2)$, where $m$ is the bond dimension of the MPS and $d$ is the number of states on one site.
Somewhat more expensive is a calculation of $E_{s}^L$, which scales as $\mathcal{O}(m^2dw + md^2w^2)$, where $w$ is  the bond dimension of the Hamiltonian MPO.  
This is a factor of the bond dimension $m$ less expensive than a DMRG sweep, which scales as $\mathcal{O}(m^3dw + m^2d^2w^2)$. 
Estimating the variance via sampling can cost less than or be comparable to a sweep, provided that we take a modest number of samples, i.e., $N_s < m$, where typically $m \sim 1000-10000$. 
In Fig.~\ref{fig03}(a) we plot both the two-site and sampled variance of an approximate solution from DMRG, for a nearest-neighbor 2D Heisenberg cluster with periodic boundary conditions. 
With on the order of a thousand samples, our method provides a better variance estimate than the two-site variance, which is quite inaccurate due to the 2D geometry and periodic boundary conditions.  

Despite its relative success in Fig.~\ref{fig03}(a), the sampled variance is noisy, with large fluctuations even after hundreds of thousands of samples; near sample 200,000 is a large jump in the running average, due to a sample with a very large local energy deviation.  
We believe the large fluctuations we see in sampling the energy variance are closely tied to the unusual energy spectrum.  
Because of these large fluctuations, 
even in cases such as Fig.~\ref{fig03}(a) where the sampled variance is more accurate, the non-stochastic nature of the two-site variance makes it a better error measure for extrapolation.

For small systems such as the one used for Fig.~\ref{fig03}(a), it is possible to enumerate every sample and compute its local energy,  
as shown in Fig.~\ref{fig03}(b) and (c).  
We see a dramatic difference in the local energy probability distributions.  
The DMRG distribution has samples with very small probability but substantial contributions to the variance. 
Consider the two samples near $P_s \sim 10^{-8}$ in Fig.~\ref{fig03}(b): these contributed $\sim 10^{-4}$ to the variance each, so they must have $|E^L_s-E|^2 \sim 10^4$.  
Such extreme outliers for the DMRG state, as shown  in Fig.~\ref{fig03}(c), explain the sudden spikes in the running average.

\emph{Biased variance extrapolation}---
To control the fluctuations,
we can bound the local energy to within a symmetric interval of the exact energy, $E$, which is known from the DMRG sweep.  Letting $\epsilon = E_s^L - E$, we define the piece-wise function
\begin{equation}\label{eq:20}
f(\epsilon)=
    \begin{cases}
        -\epsilon_{\text{max}} & \text{ if  }  \epsilon < -\epsilon_{\text{max}}\\
        \epsilon & \text{ if  }   -\epsilon_{\text{max}} \leq \epsilon \leq \epsilon_{\text{max}} \\
        \epsilon_{\text{max}} &  \text{ if  }    \epsilon > \epsilon_{\text{max}}
    \end{cases}
\end{equation}
where $\epsilon_{\text{max}} > 0$ is some specified maximum local deviation. We replace $\epsilon$ by $f(\epsilon)$ for estimating the variance. 
This transformation of the data pushes in the long tails of the the local energy distribution to $\pm \epsilon_{\text{max}}$, while leaving samples with smaller deviations untouched.  

\begin{figure}[h!]
\includegraphics[width=\linewidth]{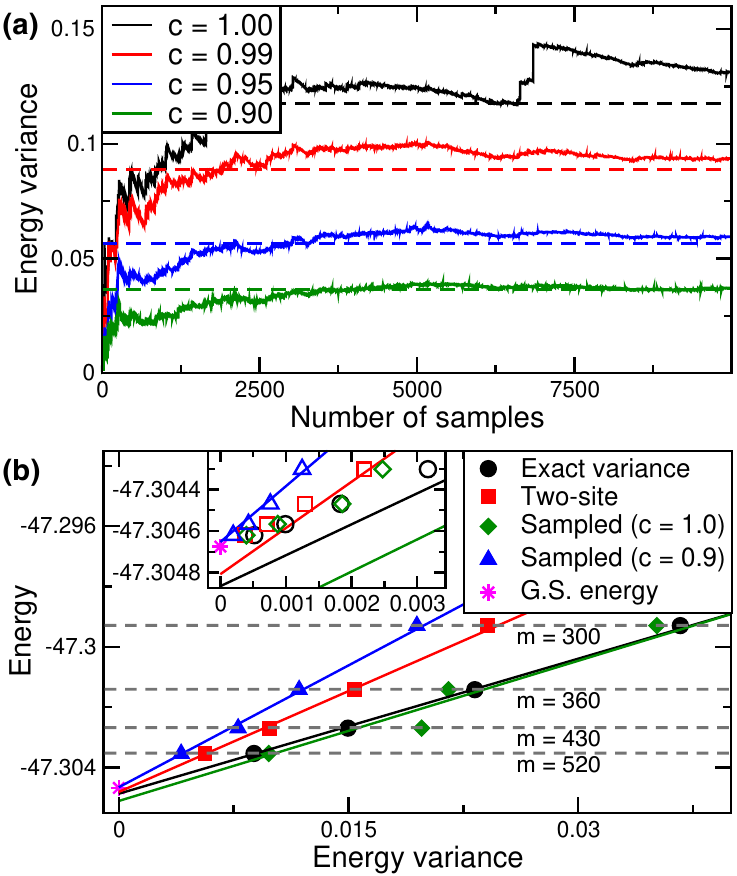}
\caption{\label{fig04} (a) For the same $4\times4$ torus, the sampled variance of a DMRG state (m = 100) at different cutoff ratios, along with exact asymptotic values. The distance from the black dashed line indicated the cutoff bias, $\Delta_c$, at each value of $c$.  (b) For a $S = 1/2$ Heisenberg model on a $12\times6$ cylinder, extrapolations of ground-state energy using the energy variance, two-site variance, and the sampled variance with and without bounding extreme samples ($N_s = 2000$ samples per bond dimension).  The unfilled data points on the inset (m = 600 - 1300) are not used for extrapolation.   }
\end{figure}

The bias induced by the cutoff, $\Delta_c$, is quite substantial; 
see Fig.~\ref{fig04}(a). 
However, in the extrapolation of the energy to zero variance as the bond dimension is increased, $\Delta_c$ decreases smoothly with the variance, and we find that this bias does not interfere with extrapolation. In order to extrapolate data, we must have a systematic way of choosing $\epsilon_{\text{max}}$ at each bond dimension. 
Define the \emph{cutoff ratio}, $c$, as the percentage of the local energies where $f(\epsilon)=\epsilon$.  A good approach for determining $\epsilon_{\text{max}}$ comes from be fixing $c$. Then the question is: what is the optimal $c$, and how well does this method extrapolate?   

For a variety of models, geometries, and boundary conditions, we have found that setting $c = 0.90$ provides excellent results; see the SM for the details~\cite{SM}.  
As shown in Fig.~\ref{fig04}(a), as $c$ decreases the sampling fluctuations drastically decrease; by $c = 0.9$ the biased variance is determined accurately with a few thousand samples. 
For this small cluster, we can explicitly compute the asymptotic values~\cite{SM}, as indicated by the dashed lines.
In Fig.~\ref{fig04}(b) we show extrapolations of the ground-state energy for a $12 \times 6$ cylindrical Heisenberg cluster, where we compare extrapolations using the exact and two-site variances, as well as from the sampled variance ($N_s = 2000$ samples) with and without the 90-th percentile cutoff.  
The higher bond dimension data points on the inset are not used to generate the best-fit lines, but rather to test their accuracy.
We find that the $c=0.9$ extrapolations offer a significant improvement over the original $c=1$ data.  
Even more, this biased sampling usually extrapolates \emph{better} than the energy variance calculated exactly without sampling. 
See the SM for more details on the performance checks for this extrapolation technique and how it compares to two-site variance extrapolations~\cite{SM}.

\emph{Summary}---We have found that high-energy tails are a universal feature of the energy spectra of MPS approximate ground states.  We have argued that this energy structure is a general consequence of any compression of a ground state and obtained similar results for neural network states. 
When comparing a DMRG result to a result from another numerical approach, e.g., quantum Monte Carlo, the energy variance of the DMRG state will be significantly larger if the two results have similar energy errors.  An even more significant consequence are large fluctuations in sampling-based approaches. The distribution of sampled local energies have fat tails, impeding the accurate determination of the energy variance. Similar large fluctuations may appear in other approaches, such as MPS-based variational Monte Carlo. However, for DMRG, 
we found that a biased variance estimator made from bounding the fluctuations can be used to produce excellent extrapolations of the ground state energy at low cost. Importantly, the sampled variance can be computed even in the context of single-site DMRG, in the presence of long range interactions, and without careful design of the sweeping procedure, making it a powerful new tool for DMRG calculations.

\emph{Acknowledgements}---This work was supported by the National Science Foundation under DMR-2110041, as well as by the Eddleman Quantum Institute.  We would also like to thank Shengtao Jiang, Miles Stoudenmire, Uli Schollwöck, and Antoine Georges for insightful conversations.

\bibliography{ms}

\begin{thebibliography}{16}%
\makeatletter
\providecommand \@ifxundefined [1]{%
 \@ifx{#1\undefined}
}%
\providecommand \@ifnum [1]{%
 \ifnum #1\expandafter \@firstoftwo
 \else \expandafter \@secondoftwo
 \fi
}%
\providecommand \@ifx [1]{%
 \ifx #1\expandafter \@firstoftwo
 \else \expandafter \@secondoftwo
 \fi
}%
\providecommand \natexlab [1]{#1}%
\providecommand \enquote  [1]{``#1''}%
\providecommand \bibnamefont  [1]{#1}%
\providecommand \bibfnamefont [1]{#1}%
\providecommand \citenamefont [1]{#1}%
\providecommand \href@noop [0]{\@secondoftwo}%
\providecommand \href [0]{\begingroup \@sanitize@url \@href}%
\providecommand \@href[1]{\@@startlink{#1}\@@href}%
\providecommand \@@href[1]{\endgroup#1\@@endlink}%
\providecommand \@sanitize@url [0]{\catcode `\\12\catcode `\$12\catcode `\&12\catcode `\#12\catcode `\^12\catcode `\_12\catcode `\%12\relax}%
\providecommand \@@startlink[1]{}%
\providecommand \@@endlink[0]{}%
\providecommand \url  [0]{\begingroup\@sanitize@url \@url }%
\providecommand \@url [1]{\endgroup\@href {#1}{\urlprefix }}%
\providecommand \urlprefix  [0]{URL }%
\providecommand \Eprint [0]{\href }%
\providecommand \doibase [0]{https://doi.org/}%
\providecommand \selectlanguage [0]{\@gobble}%
\providecommand \bibinfo  [0]{\@secondoftwo}%
\providecommand \bibfield  [0]{\@secondoftwo}%
\providecommand \translation [1]{[#1]}%
\providecommand \BibitemOpen [0]{}%
\providecommand \bibitemStop [0]{}%
\providecommand \bibitemNoStop [0]{.\EOS\space}%
\providecommand \EOS [0]{\spacefactor3000\relax}%
\providecommand \BibitemShut  [1]{\csname bibitem#1\endcsname}%
\let\auto@bib@innerbib\@empty
\bibitem [{\citenamefont {Becca}\ and\ \citenamefont {Sorella}(2017)}]{Becca_Sorella_2017}%
  \BibitemOpen
  \bibfield  {author} {\bibinfo {author} {\bibfnamefont {F.}~\bibnamefont {Becca}}\ and\ \bibinfo {author} {\bibfnamefont {S.}~\bibnamefont {Sorella}},\ }\href@noop {} {\emph {\bibinfo {title} {Quantum Monte Carlo Approaches for Correlated Systems}}}\ (\bibinfo  {publisher} {Cambridge University Press},\ \bibinfo {year} {2017})\BibitemShut {NoStop}%
\bibitem [{\citenamefont {Shi}\ \emph {et~al.}(2018)\citenamefont {Shi}, \citenamefont {Demler},\ and\ \citenamefont {Cirac}}]{shi2018variational}%
  \BibitemOpen
  \bibfield  {author} {\bibinfo {author} {\bibfnamefont {T.}~\bibnamefont {Shi}}, \bibinfo {author} {\bibfnamefont {E.}~\bibnamefont {Demler}},\ and\ \bibinfo {author} {\bibfnamefont {J.~I.}\ \bibnamefont {Cirac}},\ }\href@noop {} {\bibfield  {journal} {\bibinfo  {journal} {Ann. Phys.}\ }\textbf {\bibinfo {volume} {390}},\ \bibinfo {pages} {245} (\bibinfo {year} {2018})}\BibitemShut {NoStop}%
\bibitem [{\citenamefont {White}(1992)}]{white1992density}%
  \BibitemOpen
  \bibfield  {author} {\bibinfo {author} {\bibfnamefont {S.~R.}\ \bibnamefont {White}},\ }\href@noop {} {\bibfield  {journal} {\bibinfo  {journal} {Phys. Rev. Lett.}\ }\textbf {\bibinfo {volume} {69}},\ \bibinfo {pages} {2863} (\bibinfo {year} {1992})}\BibitemShut {NoStop}%
\bibitem [{\citenamefont {Schollw{\"o}ck}(2011)}]{schollwock2011density}%
  \BibitemOpen
  \bibfield  {author} {\bibinfo {author} {\bibfnamefont {U.}~\bibnamefont {Schollw{\"o}ck}},\ }\href@noop {} {\bibfield  {journal} {\bibinfo  {journal} {Ann. Phys.}\ }\textbf {\bibinfo {volume} {326}},\ \bibinfo {pages} {96} (\bibinfo {year} {2011})}\BibitemShut {NoStop}%
\bibitem [{\citenamefont {Stoudenmire}\ and\ \citenamefont {White}(2012)}]{Stoudenmire_2012}%
  \BibitemOpen
  \bibfield  {author} {\bibinfo {author} {\bibfnamefont {E.}~\bibnamefont {Stoudenmire}}\ and\ \bibinfo {author} {\bibfnamefont {S.~R.}\ \bibnamefont {White}},\ }\href {https://doi.org/10.1146/annurev-conmatphys-020911-125018} {\bibfield  {journal} {\bibinfo  {journal} {Annu. Rev. Condens. Matter Phys.}\ }\textbf {\bibinfo {volume} {3}},\ \bibinfo {pages} {111–128} (\bibinfo {year} {2012})}\BibitemShut {NoStop}%
\bibitem [{\citenamefont {White}\ and\ \citenamefont {Scalapino}(2009)}]{PhysRevB.79.220504}%
  \BibitemOpen
  \bibfield  {author} {\bibinfo {author} {\bibfnamefont {S.~R.}\ \bibnamefont {White}}\ and\ \bibinfo {author} {\bibfnamefont {D.~J.}\ \bibnamefont {Scalapino}},\ }\href {https://doi.org/10.1103/PhysRevB.79.220504} {\bibfield  {journal} {\bibinfo  {journal} {Phys. Rev. B}\ }\textbf {\bibinfo {volume} {79}},\ \bibinfo {pages} {220504} (\bibinfo {year} {2009})}\BibitemShut {NoStop}%
\bibitem [{\citenamefont {Depenbrock}\ \emph {et~al.}(2012)\citenamefont {Depenbrock}, \citenamefont {McCulloch},\ and\ \citenamefont {Schollw\"ock}}]{PhysRevLett.109.067201}%
  \BibitemOpen
  \bibfield  {author} {\bibinfo {author} {\bibfnamefont {S.}~\bibnamefont {Depenbrock}}, \bibinfo {author} {\bibfnamefont {I.~P.}\ \bibnamefont {McCulloch}},\ and\ \bibinfo {author} {\bibfnamefont {U.}~\bibnamefont {Schollw\"ock}},\ }\href {https://doi.org/10.1103/PhysRevLett.109.067201} {\bibfield  {journal} {\bibinfo  {journal} {Phys. Rev. Lett.}\ }\textbf {\bibinfo {volume} {109}},\ \bibinfo {pages} {067201} (\bibinfo {year} {2012})}\BibitemShut {NoStop}%
\bibitem [{\citenamefont {S.~Yan}\ and\ \citenamefont {White}(2011)}]{doi:10.1126/science.1201080}%
  \BibitemOpen
  \bibfield  {author} {\bibinfo {author} {\bibfnamefont {D.~A.~H.}\ \bibnamefont {S.~Yan}}\ and\ \bibinfo {author} {\bibfnamefont {S.~R.}\ \bibnamefont {White}},\ }\href {https://doi.org/10.1126/science.1201080} {\bibfield  {journal} {\bibinfo  {journal} {Science}\ }\textbf {\bibinfo {volume} {332}},\ \bibinfo {pages} {1173} (\bibinfo {year} {2011})}\BibitemShut {NoStop}%
\bibitem [{SM()}]{SM}%
  \BibitemOpen
  \href@noop {} {}\bibinfo {note} {See Supplemental Material at URL-will-be-inserted-by-publisher for description of lattice models, details about alternative sampling setups, analysis of energy spectra of larger systems, and performance checks of our new extrapolation method on a variety of systems.}\BibitemShut {Stop}%
\bibitem [{\citenamefont {Carleo}\ and\ \citenamefont {Troyer}(2017)}]{carleo_solving_2017}%
  \BibitemOpen
  \bibfield  {author} {\bibinfo {author} {\bibfnamefont {G.}~\bibnamefont {Carleo}}\ and\ \bibinfo {author} {\bibfnamefont {M.}~\bibnamefont {Troyer}},\ }\bibfield  {title} {\bibinfo {title} {Solving the quantum many-body problem with artificial neural networks},\ }\href {https://doi.org/10.1126/science.aag2302} {\bibfield  {journal} {\bibinfo  {journal} {Science}\ }\textbf {\bibinfo {volume} {355}},\ \bibinfo {pages} {602} (\bibinfo {year} {2017})}\BibitemShut {NoStop}%
\bibitem [{\citenamefont {Wu}\ \emph {et~al.}(2024)\citenamefont {Wu}, \citenamefont {Rossi}, \citenamefont {Vicentini}, \citenamefont {Astrakhantsev}, \citenamefont {Becca}, \citenamefont {Cao}, \citenamefont {Carrasquilla}, \citenamefont {Ferrari}, \citenamefont {Georges}, \citenamefont {Hibat-Allah} \emph {et~al.}}]{varbench}%
  \BibitemOpen
  \bibfield  {author} {\bibinfo {author} {\bibfnamefont {D.}~\bibnamefont {Wu}}, \bibinfo {author} {\bibfnamefont {R.}~\bibnamefont {Rossi}}, \bibinfo {author} {\bibfnamefont {F.}~\bibnamefont {Vicentini}}, \bibinfo {author} {\bibfnamefont {N.}~\bibnamefont {Astrakhantsev}}, \bibinfo {author} {\bibfnamefont {F.}~\bibnamefont {Becca}}, \bibinfo {author} {\bibfnamefont {X.}~\bibnamefont {Cao}}, \bibinfo {author} {\bibfnamefont {J.}~\bibnamefont {Carrasquilla}}, \bibinfo {author} {\bibfnamefont {F.}~\bibnamefont {Ferrari}}, \bibinfo {author} {\bibfnamefont {A.}~\bibnamefont {Georges}}, \bibinfo {author} {\bibfnamefont {M.}~\bibnamefont {Hibat-Allah}}, \emph {et~al.},\ }\bibfield  {title} {\bibinfo {title} {Variational benchmarks for quantum many-body problems},\ }\href@noop {} {\bibfield  {journal} {\bibinfo  {journal} {Science}\ }\textbf {\bibinfo {volume} {386}},\ \bibinfo {pages} {296} (\bibinfo {year} {2024})}\BibitemShut {NoStop}%
\bibitem [{\citenamefont {Hubig}\ \emph {et~al.}(2018)\citenamefont {Hubig}, \citenamefont {Haegeman},\ and\ \citenamefont {Schollw\"ock}}]{TwoSiteVar}%
  \BibitemOpen
  \bibfield  {author} {\bibinfo {author} {\bibfnamefont {C.}~\bibnamefont {Hubig}}, \bibinfo {author} {\bibfnamefont {J.}~\bibnamefont {Haegeman}},\ and\ \bibinfo {author} {\bibfnamefont {U.}~\bibnamefont {Schollw\"ock}},\ }\href {https://doi.org/10.1103/PhysRevB.97.045125} {\bibfield  {journal} {\bibinfo  {journal} {Phys. Rev. B}\ }\textbf {\bibinfo {volume} {97}},\ \bibinfo {pages} {045125} (\bibinfo {year} {2018})}\BibitemShut {NoStop}%
\bibitem [{\citenamefont {Stoudenmire}\ and\ \citenamefont {White}(2010)}]{MilesSampling}%
  \BibitemOpen
  \bibfield  {author} {\bibinfo {author} {\bibfnamefont {E.~M.}\ \bibnamefont {Stoudenmire}}\ and\ \bibinfo {author} {\bibfnamefont {S.~R.}\ \bibnamefont {White}},\ }\href {https://doi.org/10.1088/1367-2630/12/5/055026} {\bibfield  {journal} {\bibinfo  {journal} {New J. Phys.}\ }\textbf {\bibinfo {volume} {12}},\ \bibinfo {pages} {055026} (\bibinfo {year} {2010})}\BibitemShut {NoStop}%
\bibitem [{\citenamefont {Ferris}\ and\ \citenamefont {Vidal}(2012)}]{VidalSampling}%
  \BibitemOpen
  \bibfield  {author} {\bibinfo {author} {\bibfnamefont {A.~J.}\ \bibnamefont {Ferris}}\ and\ \bibinfo {author} {\bibfnamefont {G.}~\bibnamefont {Vidal}},\ }\href {https://doi.org/10.1103/PhysRevB.85.165146} {\bibfield  {journal} {\bibinfo  {journal} {Phys. Rev. B}\ }\textbf {\bibinfo {volume} {85}},\ \bibinfo {pages} {165146} (\bibinfo {year} {2012})}\BibitemShut {NoStop}%
\bibitem [{\citenamefont {Sandvik}\ and\ \citenamefont {Vidal}(2007)}]{sandvik2007variational}%
  \BibitemOpen
  \bibfield  {author} {\bibinfo {author} {\bibfnamefont {A.~W.}\ \bibnamefont {Sandvik}}\ and\ \bibinfo {author} {\bibfnamefont {G.}~\bibnamefont {Vidal}},\ }\href@noop {} {\bibfield  {journal} {\bibinfo  {journal} {Phys. Rev. Lett.}\ }\textbf {\bibinfo {volume} {99}},\ \bibinfo {pages} {220602} (\bibinfo {year} {2007})}\BibitemShut {NoStop}%
\bibitem [{\citenamefont {Wouters}\ \emph {et~al.}(2014)\citenamefont {Wouters}, \citenamefont {Verstichel}, \citenamefont {Van~Neck},\ and\ \citenamefont {Chan}}]{wouters2014projector}%
  \BibitemOpen
  \bibfield  {author} {\bibinfo {author} {\bibfnamefont {S.}~\bibnamefont {Wouters}}, \bibinfo {author} {\bibfnamefont {B.}~\bibnamefont {Verstichel}}, \bibinfo {author} {\bibfnamefont {D.}~\bibnamefont {Van~Neck}},\ and\ \bibinfo {author} {\bibfnamefont {G.~K.-L.}\ \bibnamefont {Chan}},\ }\bibfield  {title} {\bibinfo {title} {Projector quantum monte carlo with matrix product states},\ }\href@noop {} {\bibfield  {journal} {\bibinfo  {journal} {Phys. Rev. B}\ }\textbf {\bibinfo {volume} {90}},\ \bibinfo {pages} {045104} (\bibinfo {year} {2014})}\BibitemShut {NoStop}%
\end{thebibliography}%

\end{document}


\preprint{DMRG_Sampled_Variance}

\title{Supplementary materials: Unusual energy spectra of matrix product states   }

\author{J. Maxwell Silvester\textsuperscript{1}}
\author{Giuseppe Carleo\textsuperscript{2}}
\author{Steven R. White\textsuperscript{1}}
\affiliation{
    \textsuperscript{1}Department of Physics and Astronomy, University of California, Irvine, California 92667, USA
}
\affiliation{
    \textsuperscript{2}Institute of Physics, École Polytechnique Fédérale de Lausanne (EPFL), CH-1015 Lausanne, Switzerland
}

\date{\today}
\maketitle

\subsection{Biased vs unbiased sampling}
\begin{figure}[h!]
\includegraphics[width=\linewidth]{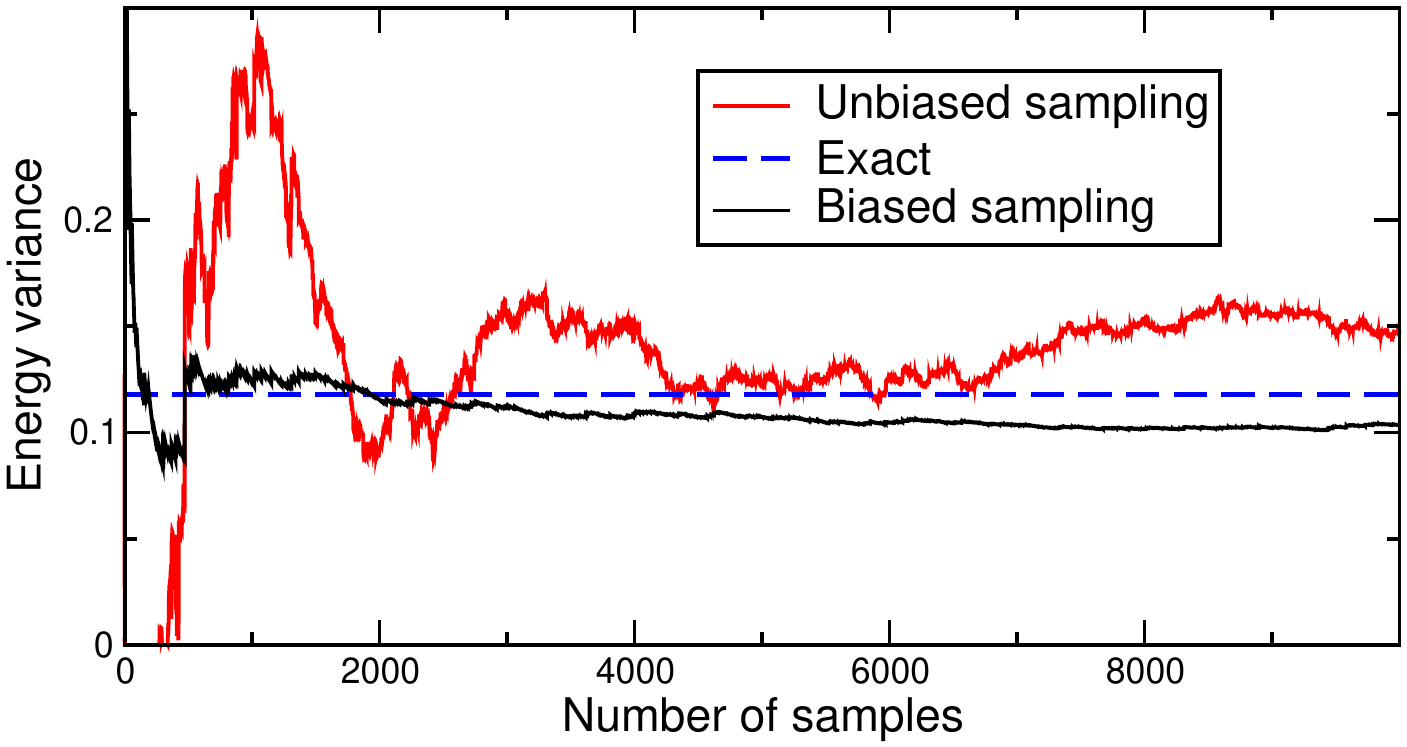}
\caption{\label{fig07} Running average of the sampled energy variance of a DMRG wavefunction with bond dimension $m=100$, for the $4 \times 4$ Heisenberg torus. The black curve uses sampled local energies (squared), i.e., Eq.(7) of the main text.  The red curve uses sampled local variances, i.e., Eq.(3) of the SI.  The blue dashed line shows the exact energy variance of the state.}
\end{figure}

To form an unbiased approximation of the energy variance we can sample the \emph{local variance}, 
\begin{equation}\label{eq:8}
\sigma^L_s = \frac{\braket{s| (\hat{H} - E)^2 | \psi}}{\braket{s|\psi}}.
\end{equation}
Just as with the local energies, the local variances are only defined on the wavefunction's support.  The energy variance is
\begin{equation}\label{eq:6}
\sigma^2_{\hat{H}} = \sum_{s
}  \braket{\psi | s} \braket{s| (\hat{H} - E)^2 | \psi}.
\end{equation}
Note that any state that is not in the support, $\mathcal{S}$, necessarily contributes zero to this sum, since its overlap with the wavefunction is exactly zero, by definition.  We can therefore replace the sum over all basis elements in Eq.~(\ref{eq:6}) with a sum restricted to the support to get
\begin{equation}\label{eq:7}
\sigma^2_{\hat{H}} = \sum_{\ket{s} \in \mathcal{S}
}  P_s  \sigma^L_s,
\end{equation}
where $ P_s = |\braket{\psi | s}|^2$.  While this sampling is unbiased, as shown in figure Fig.~\ref{fig07}, the results  are much noisier than using the biased local energy sampling.  Moreover, local variance samples are about a factor of $w$ more expensive than local energy samples, where $w$ is the bond dimension of the MPO.   

Another way to achieve unbiased sampling, which works for spin systems, is to sample in a basis different from that used for the MPS calculation, for example, sampling spins with measurements in the $x$ direction instead of the usual $z$ direction.  With this approach, no product state will have exactly zero weight. However, while this basis change technically eliminates the local energy bias, in practice it only hides it in product states with extremely small probabilities but which contribute a finite amount to the variance.  For a spin-$1/2$ system, this rotated-basis sampling is about a factor of four more expensive than the $z$-basis sampling, so again the naive approach is simpler and cheaper. 

\begin{figure}[h]
\includegraphics[width=\linewidth]{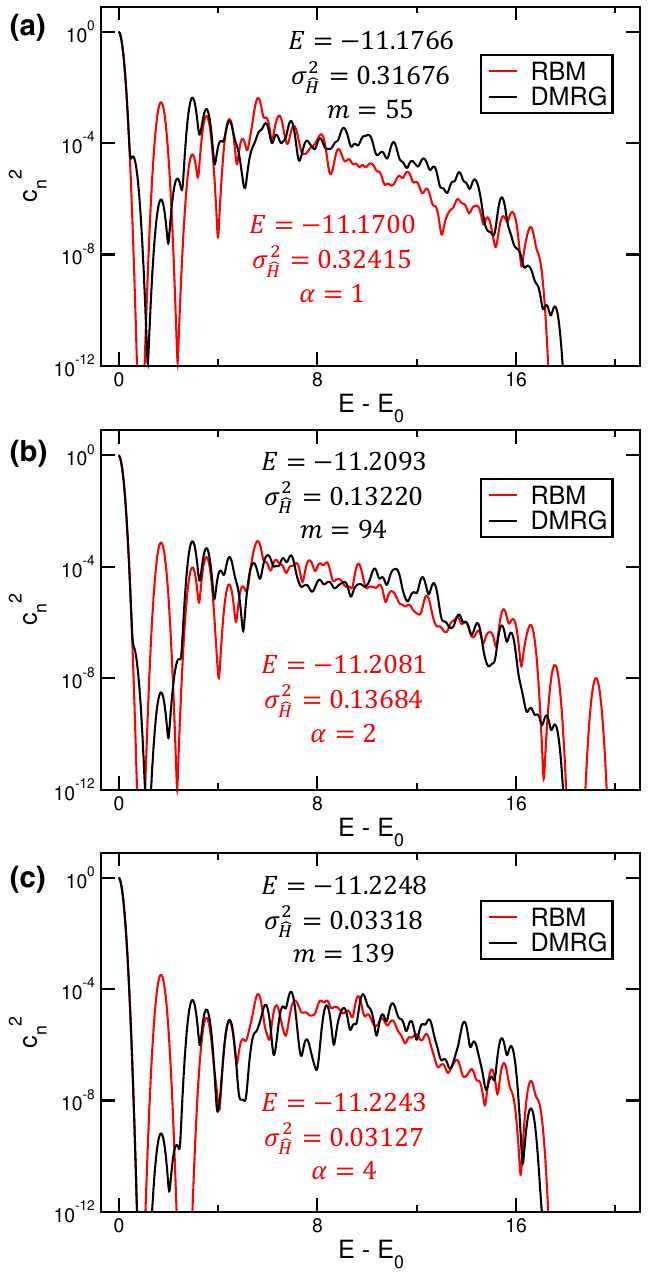}
\caption{\label{fig21} For a $4\times 4$ Heisenberg torus, comparing the energy spectra of DMRG solutions to RBM wavefunctions.  For each panel, the two states have comparable energy error and variance, as shown on the plots.  There are 18, 35, and 69 parameters for panels (a), (b), and (c), respectively.   }
\end{figure}

\subsection{Results for neural network states}
A restricted Boltzmann machine (RBM) is a simple neural network that can also be used to solve for the ground states of the lattice models described below. The RBM wavefunction with $M$ hidden units can be written as:
\begin{equation}
\Psi_{\mathrm{RBM}}(s) = e^{\sum_{i=1}^N a_i s_i} \prod_{j=1}^M \left(1 + e^{b_j + \sum_i W_{ij} s_i}\right)
\end{equation}
where $s_i$ represents the $N$ physical spins, $a_i$ are called visible biases, $b_j$ are called the hidden biases, and $W_{ij}$ are the weights connecting the visible and hidden layers. Note that given real wavefunction parameters, the wavefunction is positive for all configurations, which is valid for the near-neighbor bipartite square lattice Heisenberg model. The built-in positiveness helps reduce the number of parameters needed for high accuracy--i.e. it achieves higher compression. To reduce the number further and enforce physical symmetries, we implement translational invariance in the network architecture, constraining the weights and biases to be periodic across the lattice.

By changing the value of the hidden unit density $\alpha \equiv M/N$, we correspondingly achieve lower or higher accuracy in the RBM wavefunction. 
In Fig.~\ref{fig21}, we compare the wavefunctions from DMRG and RBM at comparable values of energy and variance. We see that qualitatively, the wavefunctions are quite similar. The one obvious discrepancy between the two is that the RBM solution has overlap with a low-lying excited state near $E-E_0=1.7$ that is orders of magnitude higher than the DMRG solution's overlap with the same state. Of course, it is not surprising that different forms of compression yield slight variations in the energy spectrum. In any case, these results support the proposition that high-energy tails are a necessary consequence of compressing the ground state. 

The number of parameters in an MPS scales as $m^2 N$, while for the RBM state the scaling is $\alpha N$ ($\alpha N^2$ if translational invariance had not been used). 
This means that for Fig.~\ref{fig21}(a) the DMRG state requires  on the order of $\sim 48400$ parameters, while the RMB state requires only $\sim 16$.  While the number of parameters required for the MPS can be reduced by enforcing symmetries, there will still be a discrepancy in the number of parameters required to achieve the same quality of solution with the two methods. Thus the number of parameters is not useful in characterizing the anomolous spectra. Instead, the results suggest that if the variance in two wavefunctions is similar, the overall features of the high energy spectrum is also likely to be similar.

\begin{figure}
\includegraphics[width=0.98\linewidth]{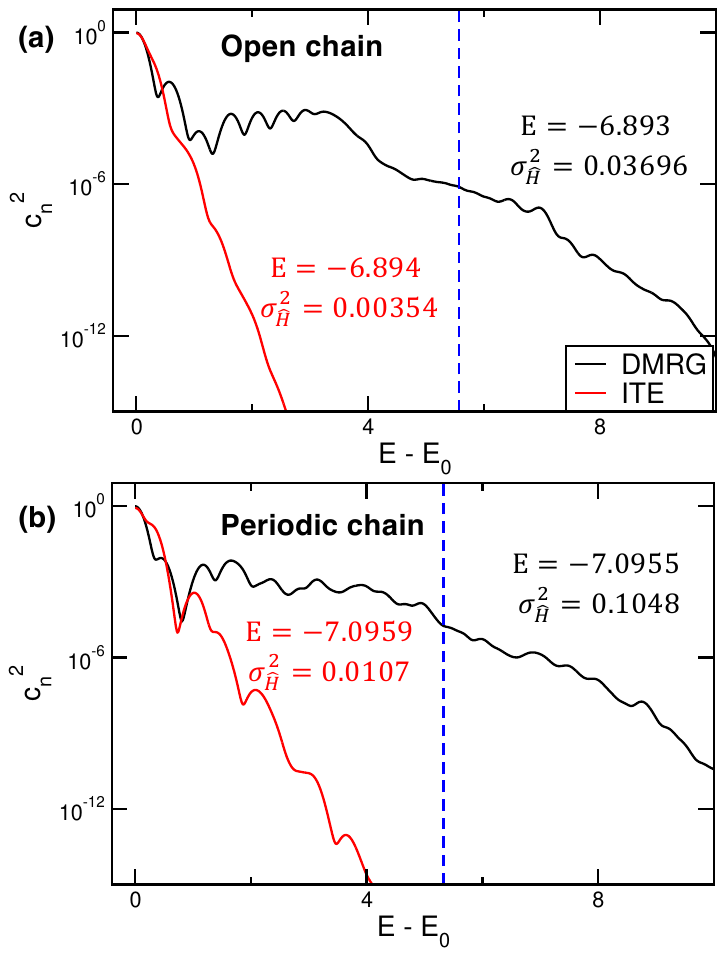}
\caption{\label{fig20} Energy spectra of both DMRG and ITE solutions for 1D Heisenberg chains. As in the main text, spectra are broadened with Gaussians and the vertical dashed line indicates half the energy bandwidth of the system. }
\end{figure}

\subsection{Results for one-dimensional chains}
One dimensional (1D) chains have less long-range entanglement than 2D geometries, as described by the area law of entanglement.  As a consequence, for a fixed number of sites, a smaller bond dimension is required for chains than for 2D systems, in order to achieve the same level of accuracy.  Therefore, since we are limited to $N\sim16$ sites for exact diagonalization,  we will only required a bond dimension $m \leq 10$ to achieve states with non-negligible energy variance for Heisenberg chains. In Fig.~\ref{fig20}, we show results for both open ($m=4$) and closed ($m=8$) chains for $N = 16$ sites.  In both cases, we see that the DMRG solution has high-energy tails similar to those observed for the 2D cases in the main text, while the ITE state's spectrum falls off exponentially, as expected.  Furthermore, while the solutions are chosen so that the energies of the DMRG and ITE states match, the energy variance of the DMRG state is an order of magnitude greater than the ITE state's.  Again, this is similiar to what we observe in the 2D geometries. These results confirm that high-energy tails are not a due to the 2D geometry, but to the compression of the state.

\subsection{Probing spectra without full diagonalization}  
\begin{figure}[!h]
\includegraphics[width=\linewidth]{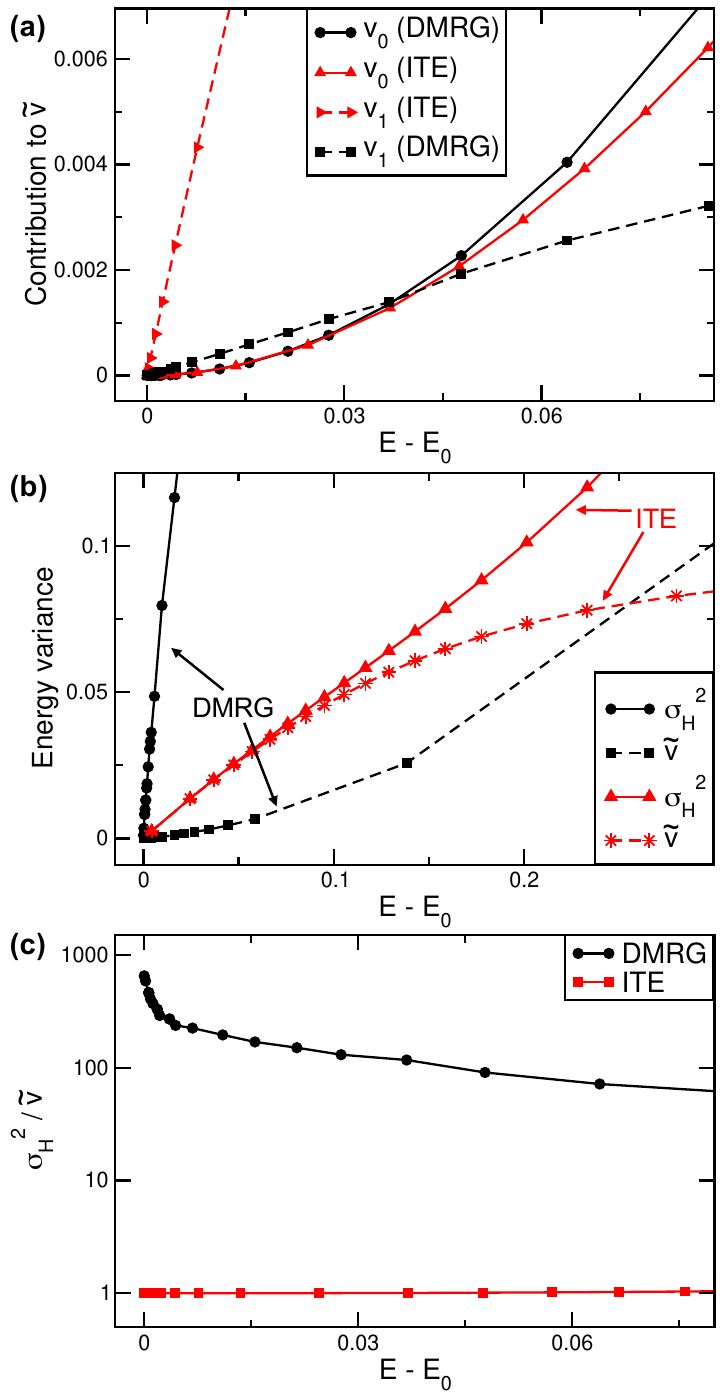}
\caption{\label{fig12}(a) Components of the approximated energy variance, $\Tilde{v}$, versus the energy error of the DMRG and ITE states, for the $4 \times 4$ Heisenberg torus. For the same system, (b) shows the exact versus approximated energy variance and (c) shows the ratio $\sigma_{\hat{H}}^2 / \Tilde{v}$. }
\end{figure}

Exact diagonalization and exact ITE without truncation quickly become infeasible as the system size increases. To confirm that the unusual energy structure of MPS wavefunctions holds generically over a broad range of models, boundary conditions and geometries, we require another approach.  Here we show that the energy variance helps us to indirectly---yet reliably---measure the extent of high-energy contributions to an approximate ground state.  First, note that the energy variance can be rewritten as
\begin{equation}\label{eq:60}
\begin{aligned}
\sigma^2_{\hat{H}} & = \braket{\psi|(\hat{H} - E)^2|\psi} = \sum_{\ket{n}} \braket{\psi|n} \braket{n| (\hat{H} - E)^2|\psi} \\
& = c_0^2 (E - E_0)^2 + \sum_{n \neq 0} c_n^2 (E - E_n)^2, 
\end{aligned}
\end{equation}
where $\ket{n}$ are again the energy eigenstates.  For a well-converged ITE state, we know that the sum over excited states on the right-hand side of Eq.~(\ref{eq:60}) will only have non-negligible contributions from eigenstates with energy near $E_1$; the coefficients $c_n^2$ will be exponentially small if $E_n \gg E_1$.  Therefore, for ITE states we may approximate the energy variance by replacing $E_n$ with $E_1$ in  Eq.~(\ref{eq:60}), reaching
\begin{equation}\label{eq:61}
\begin{aligned}
\sigma^2_{\hat{H}} & \approx c_0^2 (E - E_0)^2 + \sum_{n \neq 0} c_n^2 (E - E_1)^2  \\
 & = c_0^2 (E - E_0)^2 + (1 - c_0^2) (E - E_1)^2 = \Tilde{v}.
\end{aligned}
\end{equation}
In Fig.~\ref{fig12}(a), we show, again for the same $4 \times 4$ Heisenberg torus studied in the main text, how $v_0 = c_0^2 (E - E_0)^2$ and $v_1 = (1 - c_0^2) (E - E_1)^2$ contribute to $\Tilde{v}$ as the energy error decreases (by increasing the bond dimension for the DMRG state and increasing imaginary-time for the ITE state). Note that $v_0$ is an exact term contributing to the variance, while $v_1$ contains all of the approximations that allows $\Tilde{v} = v_0 + v_1$ to deviate from $\sigma_{\hat{H}}^2$.

While we expect that $\Tilde{v}$ gives a good estimate of the variance for the ITE state, we do \emph{not} expect this to be the case for the DMRG state; this is because our assumption about the exponential fall-off of the coefficients $c_n^2$ is not valid in this latter case.  In Fig.~\ref{fig12}(b) we compare $\Tilde{v}$ to the exact variance for both the DMRG and ITE states.  As the energy error decreases, the two values match more and more closely for ITE state.  However, $\Tilde{v}$ severely underestimates the variance for the DMRG state, even at small energy errors.

The ratio $\sigma^2_{\hat{H}}/\Tilde{v}$ therefore offers us important information about the energy spectrum.  If the ratio is near unity, then the residue is dominated by low-lying excited states; alternatively, if $\sigma^2_{\hat{H}}/\Tilde{v} \gg 1$, then there are significant high-energy contributions to the residue.  As shown in Fig.~\ref{fig12}(c), the ratio approaches unity as the ITE state converges, while for the DMRG state the ratio \emph{increases} with decreasing energy error.  This result for the DMRG state is a consequence of the fact that average energy of the residue increases with increasing bond dimension, as explored in the main text;  the increasing energy of the residue makes the assumption about the coefficients $c_n^2$ less accurate as the bond dimension increases and the approximation in Eq.~(\ref{eq:61}) grows correspondingly worse.

In order to calculate $\sigma^2_{\hat{H}}/\Tilde{v} = 1$ we require four quantities:  the energy of the wavefunction (which comes for free during a DMRG sweep), the overlap of the wavefunction with the exact ground state, and the energies of both the ground- and first excited-state.  For larger systems, we cannot compute $c_0$, $E_0$ and $E_1$ via exact diagonalization, as we did for the small $4 \times 4$ cluster above.  Instead, we rely on highly converged DMRG results ($m \approx 10000$) to estimate the ground state and first excited state.  

\subsection{Spectra for various lattice models}
In this section we provide a brief description of the lattice models and geometries that we study.  We then provide results indicating similar high-energy tails for all DMRG solutions, no matter the specifics of the geometry or model.  
\subsubsection{$S = 1/2$ Heisenberg model}
The $S = 1/2$ Heisenberg model is a nearest-neighbor quantum spin model, useful in the study of a broad range of magnetic phenomena, including antiferromagnetism, ferromagnetism, and spin liquids.  The model is described by the Hamiltonian
\begin{equation}\label{eq:15}
H = J \sum_{\braket{i,j}} \vec{S}_i \cdot \vec{S}_j,
\end{equation}
where the sum ranges over nearest-neighbor pairs.  
We set $J = 1$ throughout, corresponding to antiferromagnic coupling, considering both square and triangular lattices. In the main text we present results for $4\times4$ square lattice clusters with various boundary conditions.  For the study of larger clusters below, we also perform calculations for a $12\times6$ square-lattice cylinder. We also consider a triangular lattice, for which the orientation matters when considering periodic boundary conditions. In our calculations, the lattice is oriented so that when we impose cylindrical boundary conditions, a third of nearest-neighbor bonds are parallel to the periodic direction, with no bonds parallel to the open direction. To avoid frustration, we require a multiple of three sites along this periodic direction. We choose a $6\times6$ triangular lattice cylinder to study in the following.

\subsubsection{$J_1$-$J_2$ model}
The $J_1$-$J_2$  generalizes the  Heisenberg model on a square lattice to include next-nearest-neighbor interactions, with Hamiltonian
\begin{equation}\label{eq:19}
H = J_1 \sum_{\braket{i,j}} \sigma_i \sigma_j + J_2\sum_{\braket{\braket{i,j}}}\sigma_i \sigma_j,
\end{equation}
where the first and second sums over over nearest- and next-nearest-neighbors, respectively.  In this work, we set $J_1 = 1$, which fixes the energy scale.  We then choose $J_2 = 0.2$, a relatively strong next-nearest neighbor coupling, so that the system is clearly distinguish from a pure Heisenberg interaction. In the following, we present results for an $8\times8$ open cluster.

\subsubsection{Hubbard model}
The Hubbard Hamiltonian is
\begin{equation}\label{eq:22}
H = -t \sum_{\braket{i,j},s} (c_{i,s}^\dagger c_{j,s} + c_{j,s}^\dagger c_{i,s}) +   U \sum_i n_{i \uparrow} n_{i \downarrow},
\end{equation}
where  $c_{i,s}^\dagger$ and $c_{i,s}$ are the creation and annihilation operators for an electron on site $i$ with spin $s$, and $n_{i,s} = c_{i,s}^\dagger c_{i,s}$.
The sum in the first term is over nearest-neighbor pairs $\braket{i,j}$ and spins $s \in \{\uparrow, \downarrow\}$. We set $t=1$ and $U=6$ below. We consider only square lattices at half-filling, with an average of one particle per site, focusing on a $8\times4$ cylinder.

\begin{figure}[h!]
\includegraphics[width=0.98\linewidth]{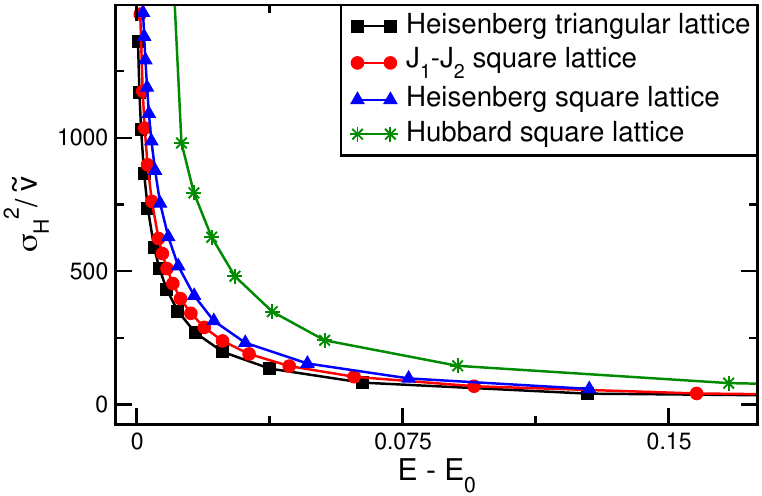}
\caption{\label{fig09}  The ratio  $\sigma^2_{\hat{H}}/\Tilde{v}$ as DMRG states converge, demonstrating similar high-energy contributions to approximate ground states of larger systems: a $12\times6$ Heisenberg square lattice with cylindrical boundary (blue); a $6\times6$ Heisenberg triangular lattice with cylindrical boundary (black); an $8\times 8$ $J_1$-$J_2$ square lattice with open boundary and $J_2 = 0.2$ (red); and an $8\times4$ Hubbard square lattice with cylindrical boundary and $U/t = 6$ (green).  In all cases, $\Tilde{v}$ is computed by finding highly-converged estimates of the ground and first-excited states via DMRG.   }
\end{figure}

\subsubsection{Results}
For small clusters like the one explored in the main text, we find similar high-energy tails across these different lattice models, in all cases where the full diagonalization calculations can be performed.  These investigations include a variety of geometries and boundary conditions.  Of more interest is whether similar spectra arise from MPS representation of larger clusters, where DMRG is of practical use since exact diagonalization is not available.  We therefore rely on the ratio $\sigma^2_{\hat{H}}/\Tilde{v}$ described in the previous section as a signal of high-energy tails.  Results from such analysis are shown in Fig.~\ref{fig09}.  We see that in all cases the ratio grows rapidly as the wavefunction converges to zero energy error, confirming that these MPS solutions have a similar energy structure to the smaller clusters explored in the main text.

\subsection{Computing the cutoff bias} 
In the continuum limit, we define the density of states, $P(E_L - E)$, centered around zero and normalized so that 
\begin{equation}\label{eq:10}
\int_{-\infty}^{+\infty} P(E_L - E)\,dE_L = 1.
\end{equation}
Recall that the local energies and therefore $P(E_L - E)$ are only defined over the support of the wavefunction.  With the local energy bias in mind, we may write
\begin{equation}\label{eq:9}
\sigma_{\hat{H}}^2 - \Delta_s = \int_{-\infty}^{+\infty} P(\epsilon)\epsilon^2 \,d\epsilon,
\end{equation}
where we have changed variables $E_L-E\to\epsilon$.  Next, for a given cutoff ratio, $c$, we may find the corresponding cutoff energy, $\epsilon_c\ > 0$, which satisfies
\begin{equation}\label{eq:11}
c = \int_{-\epsilon_c}^{+\epsilon_c} P(\epsilon) \,d \epsilon .
\end{equation}
Note that $\epsilon_{\text{c}}$ is distinguished slightly from $\epsilon_{\text{max}}$ used in the main text, since the latter is not necessarily unique.  Specifically, there may be a finite gap between two adjacent local energy values; then, any cutoff energy in this gap will yield the same cutoff ratio.  On the other hand, there will be exactly one value of $\epsilon_c$ satisfying Eq.~(\ref{eq:11}).   The bias is then
\begin{equation}\label{eq:12}
\Delta_c = \int_{-\infty}^{-\epsilon_c} P(\epsilon) (\epsilon + \epsilon_c)^2 \,d\epsilon +  \int_{+\epsilon_c}^{+\infty} P(\epsilon) (\epsilon - \epsilon_c)^2 \,d\epsilon .
\end{equation}

For a finite system, we rewrite the integrals as sums by introducing the set $\mathcal{C} \subseteq \mathcal{S}$ of basis states in the support that are cut off because of their extreme local energies.   Then, the cutoff ratio is
\begin{equation}\label{eq:13}
c = \sum_{ s \notin \mathcal{C}} P_s
\end{equation}
and the cutoff bias is
\begin{equation}\label{eq:14}
\Delta_c = \sum_{s \in \mathcal{C}} P_s (\epsilon_c - |\epsilon|) ^2 . 
\end{equation}
For a small enough cluster, we can compute the local energies of each basis state and therefore determine the set, $\mathcal{C}$, of cutoff states needed to obtain a specified ratio. Of course, for these finite systems, the cutoff ratio can only take on certain discreet values, so there will be some lack of precision in achieving a cutoff ratio of exactly $c = 0.90$.  However, this error vanishes quickly as the system size increases. For example, for the small $4 \times 4$ Heisenberg torus that we explore frequently in the main text, a DMRG solution with $\sigma_{\hat{H}}^2 = 0.1175$ and $m = 100$, has about $12868$ states in the support. This number may fluctuated slightly depending on the ``warm up'' sweeps used at smaller bond dimension. To approximate a $10\%$ cutoff, we bound the $|\mathcal{C}| = 4774$ most extreme states and obtain a cutoff ratio of $c = 0.9000100$ and cutoff energy of $\epsilon_c = 0.4824$.   The cutoff bias is then found to be $\Delta_c = 0.0809$, using Eq.~(\ref{eq:14}).

\begin{figure}[h!]
\includegraphics[width=\linewidth]{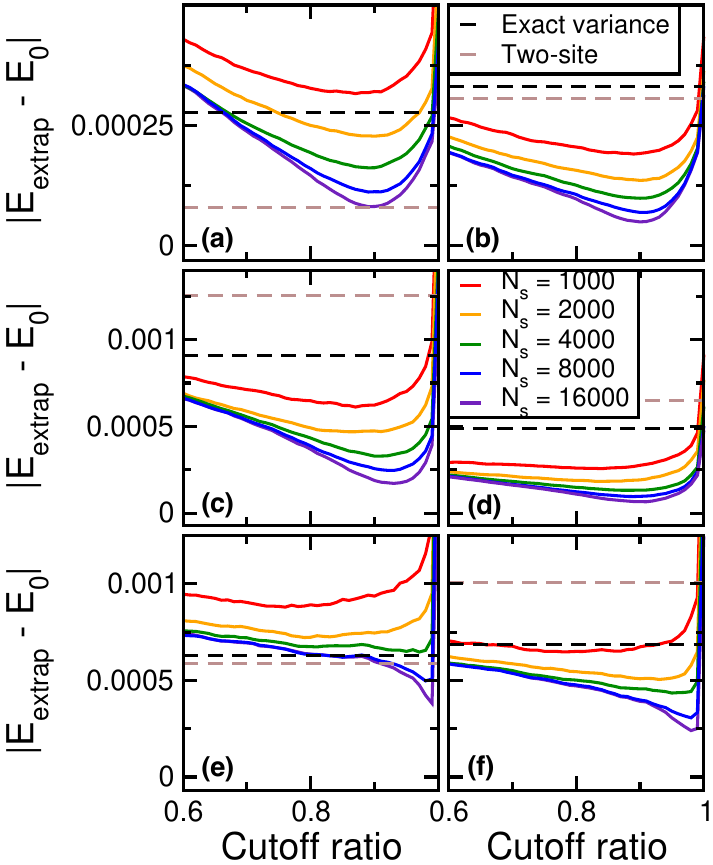}
\caption{\label{fig08}  Optimizing the cutoff ratio for biased variance extrapolations. At each value of the cutoff ratio and the number of samples, $N_s$, 1000 sampling-based extrapolations are performed and the average energy errors are plotted as solid lines. Exact and two-site variance extrapolations from the same sweeps are shown as dashed lines. (a) - (b) Heisenberg model on a $12 \times 6$ square lattice with cylindrical boundary conditions.  The maximum bond dimensions of the last sweeps are $m = 350$ and $m=450$, respectively.   (c) - (d) Heisenberg model on a $6 \times 6$  triangular lattice with cylindrical boundary conditions; $m=1000$ and $m=1400$, respectively.  (e) - (f) $8\times 4$ Hubbard cylinder; $m=1400$ and $m=1700$.   }
\end{figure}

\subsection{Extrapolation and optimizing the cutoff ratio} 

Historically, it has been standard practice in DMRG calculations to perform extrapolations in the \emph{truncation error}, i.e., the total probability of the states that are thrown away during sweeping to compress the state.
Not only do such extrapolations provide error bars, but non-linearity in the energy versus truncation error can also indicate that the initial wavefunction was far from the ground state and the system is not well converged.  Although the truncation error is estimated at almost zero computational cost during a DMRG sweep, this estimate depends on details of the sweep. Producing a reliable truncation error therefore requires care in the construction of the sweeping procedure and may be impossible for systems with long-range interactions. Moreover, estimating the truncation error is not possible in the single-site implementation of DMRG, which is cheaper than the two-site algorithm. 

These well-established truncation error extrapolations use data from successive DMRG sweeps at increasing bond dimension to create weighted linear fits.  That is, since data from higher bond dimension sweeps are more accurate, they are weighted more heavily than lower bond dimension data. Specifically, the error for each data point is chosen to be the truncation error of that point. In other words, the point $(x,y)$ is assumed to have an error bar of $x$. Throughout the Letter and this SI, we utilized this same weighted-fit technique when performing extrapolations in the energy variance and surrogates thereof, i.e., two-site variance and sampled variance.  For all extrapolations, we use data from four consecutive bond dimensions.

While the \emph{two-site variance} mentioned in the Letter is usually a very accurate approximation of the energy variance for nearest-neighbor one-dimensional open chains, it can deviate strongly for 2D systems. However, its errors are systematic, and the two-site variance vanishes when the state is an exact eigenstate. Thus, this quantity can be used to produce high-quality extrapolations. Still, it is relatively expensive to compute.  The drawbacks of both the truncation error and the two-site variance motivate us to find another error metric that is both inexpensive to compute and produces high-quality extrapolations, even when running the single-site DMRG algorithm. For this reason, we consider estimating the energy variance via sampling as described in the Letter.  

For our extrapolations from bounded local energy sampling, we wish to find a value of the cutoff ratio that can be held constant across all bond dimensions, yielding high-quality extrapolations.  For this optimization, we switch to larger systems more similar to those which would be of interest in a practical DMRG calculation.  Still, the system must still be small enough so that we can compute ground-state energies to near numerical precision, enabling us to judge the accuracy of extrapolations from lower dimension data.  

For an extrapolation at some fixed value of $c$, the absolute energy error $|E_{\text{extrap}} - E_0|$ is computed, where $E_0$ is the exact ground-state energy computed via much higher bond dimension DMRG calculations, $m \approx 10000$.  We repeat this process, generating a batch of $N_s$ samples at each bond dimension, and  computing an energy error.  Finally, we average the results from 1000 such batches.  In this way we can compute the average energy error at a particular $c$-value and compare the result to the errors obtained from extrapolations using both the exact and two-site energy variance. Results  from such an analysis for a variety of models and geometries are shown in Fig.~\ref{fig08}.  We see that the lowest average energy error from our sampling consistently occurs near $c = 0.9$.  In some cases, the absolute minimum might be closer to $c = 0.95$ or higher, such as for the Hubbard model plots.  However, in other cases a cutoff ratio greater than $0.9$ decreases the accuracy of extrapolations, as in the Heisenberg square lattice plots.  In all cases $c = 0.9$ is close to optimal and performs competitively, or simply better, than the exact or two-site extrapolations.  It is interesting to note that the two-site variance seems to yield extremely accurate extrapolations in Fig.~\ref{fig08}(a). However, performing just one more sweep at a modestly increased bond dimension, the two-site extrapolations are significantly worse the the average results from sampling with $c = 0.9$, as shown in Fig.~\ref{fig08}(b).  In general it seems that the energy error from the two-site extrapolation is anomalously low in (a).